# Variability of Interplanetary Shock and Associated Energetic Particle Properties as a Function of the Time Window Around the Shock

K. Moreland[1,2] , M. A. Dayeh[1,2] , G. Li[3] , A. Farahat[4] , R. W. Ebert[1,2] , and M. I. Desai[1,2]
[1] Southwest Research Institute, San Antonio, TX 78238, USA
[2] University of Texas at San Antonio, San Antonio, TX 78249, USA
[3] University of Alabama in Huntsville, Huntsville, AL, 35899, USA
[4] King Fahd University of Petroleum & Minerals, Dhahran, Saudi Arabia

## Abstract

We study the effect of sampling windows on derived shock and associated energetic storm particle (ESP) properties in 296 fast-forward interplanetary shocks using Advanced Composition Explorer measurements at 1 au between 1998 February and 2013 August. We vary the time windows from 2 minutes to 20 minutes for the shock properties and from 2 minutes to 540 minutes for ESP properties. Variability is quantified by the median absolute deviation statistic. We find that the magnetic, density, and temperature compression ratios vary from their median values by 17.03%, 20.05%, and 25.91%, respectively; shock speed by 16.26%; speed jump by 45.46%; Alfvénic Mach number by 31.53%; and shock obliquity by 24.25%. Spectral indices in the 2 minute–540 minute windows downstream of the shock vary from the median value of 1.79 by 26.05% and by 30.53% from the 1.70 median value upstream of the shock. Similarity of ESP spectral indices upstream and downstream of the shock suggest that these ESP populations are likely locally accelerated at the shock. Furthermore, we find that for a moving sampling window around the shock, values for the density ratio hold for ~10 minutes; the magnetic ratio and shock speed jump hold for ~30 minutes and ~60 minutes, respectively. Fixing the upstream window to 2 minutes and moving only in the downstream direction, the density ratio holds for ~60 minutes downstream, magnetic ratio holds for ~30 minutes, and the shock speed jump holds for ~110 minutes. Beyond these time windows, derived shock properties are no longer representative of shock properties. These results provide constraints for modeling and forecasting efforts of shock and ESP-associated properties.

*Unified Astronomy Thesaurus concepts:* [Interplanetary shocks (829)](#); [Solar energetic particles (1491)](#); [Solar coronal mass ejections (310)](#); [Interplanetary particle acceleration (826)](#); [Timing variation methods (1703)](#)

## 1. Introduction

Interplanetary (IP) shocks are driven by fast IP coronal mass ejections (ICMEs) and continue to accelerate particles as they propagate through space, often producing particle enhancements, known as energetic storm particle (ESP) events (e.g., Desai et al. 2003; Lario et al. 2003; Desai & Giacalone 2016; Dayeh et al. 2018). ESPs are a type of solar energetic particle (SEP) event that can bring sudden and significant increases to the near-Earth particulate radiation. This increased radiation poses serious hazards to astronauts and assets in space. High-energy particles can disrupt spacecraft electronic operations and lead to degradation of the instrumentation on board. They can also cause data disruption or biasing of instrument readings and, when strong enough, lead to irreversible damage to instruments (e.g., Horne et al. 2013; Maurer et al. 2017). The penetrating particle radiation also affects commercial aviation communication and navigation and can affect the health of the crew and passengers on polar flights. Astronauts in space can suffer acute and long-term illnesses, such as increased cancer risks (e.g., Chancellor et al. 2014; Onorato et al. 2020). The need to understand the link between IP shocks and their associated ESPs is thus an integral step in advancing modeling and predictions of the physical properties of ESPs and SEPs, such as their arrival time and peak intensities (e.g., Dayeh et al. 2010).

Acceleration of particles at shocks is dominated by the diffusive shock acceleration (DSA; e.g., Decker 1981, Lee 1983) process. DSA predicts the resulting particle spectrum to be a power-law distribution with an index that depends on key parameters such as the shock speed and the magnetic field configuration. Nonetheless, observations show large variability that is attributed to several affecting factors, including ambient plasma conditions, local shock properties, and fluctuations in the local magnetic field (e.g., Lario et al. 2005; Giacalone 2012). This work examines the effect of the time window sampling size when performing upstream and downstream IP shock and ESP studies. Quantifying this effect is essential as we utilize machine-learning algorithms for mining and automating space-weather data sets and advancing their forecasting and nowcasting capabilities.

The properties of IP shocks can vary significantly over time and space. As a result, accurate determination of shock properties requires precise and consistent measurements of the plasma and magnetic field in the vicinity of the shock. The sampling intervals, or time windows over which the data is averaged, can play a critical role in determining the accuracy and reliability of the measurements. The same applies to examining particle populations around shocks.

Over the last few decades, numerous studies on IP shocks (van Nes et al. 1984; Desai et al. 2003; Lario et al. 2003; Oh et al. 2007; Ho et al. 2008; Giacalone 2012; Reames 2012b; Kilpua et al. 2015; Dayeh et al. 2018) and their associated ESPs have used different time intervals for calculating both the IP shock and ESP properties. This was done either by manually reviewing the data and choosing windows separately for each event or by subjectively selecting a single time window for all events. In all these studies, however, selected time periods before and after the shock varied and were determined based on the quantity intended to be examined, and no sensitivity analysis was performed to see how varying the time window affects the inferred conclusions.

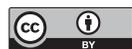

In this work, we examine the effect of the sampling window size on the shock parameters and the associated ESP properties. Determining how the time window affects the IP shock and ESP properties provides information on the ambient conditions around shocks and provides constraints for models and machine-learning automated algorithms. In particular, we quantify the variation in the derived shock parameters from 2 minutes to 20 minutes for the shock properties and 2 minutes to 540 minutes for the ESP properties. Section 2 describes the instrumentation and data sets. Section 3 explains the data selection and analysis methods. Section 4 presents the results, and Section 5 provides the conclusions.

## 2. Instrumentation

This study uses data from multiple instruments on board the Advanced Composition Explorer (ACE) spacecraft, launched in 1997 and currently in orbit near the L1 Lagrange point. We obtain 12 s energetic particle measurements for ions from the Electron, Proton, and Alpha Monitor (EPAM; Gold et al. 1998). On EPAM, the Low Energy Magnetic Spectrometer (LEMS), monitors low-energy (0.47–4.8 MeV) ions that are separated into eight channels. The LEMS120 head points 120° from the spacecraft axis, looking primarily in the anti-sunward (earthward) direction. The LEMS30 head points 30° from the spin axis, looking nearly into the Sun and observes higher intensities than LEMS120. As a result, X-ray contamination is present at almost all times in the LEMS30. Plus, the P1 and P2 channels have no data after day 371 of 2001 and after day 302 of 2003 (Haggerty et al. 2006). For this reason, we choose only to use the LEMS120 data in our study. The shock properties are analyzed using 64 s resolution solar wind plasma data from the Solar Wind Electron, Proton, and Alpha Monitor (SWEPAM; McComas et al. 1998). Magnetic field data at a resolution of 16 s from the Magnetic Field Experiment (MAG; Smith et al. 1998) is used. We also utilize the Helsinki Database of Heliospheric Shocks Waves, generated and maintained by the University of Helsinki.[5]

## 3. Data Selection and Analysis Methods

### 3.1. Data and Shock Selection

We begin the analysis using 296 fast-forward shocks detected by the ACE spacecraft at 1 au from 1998 to 2013 during solar cycles 23 and 24. The shocks are identified in the Helsinki database using the following selection criteria: speed jump > 20 km s$^{-1}$, magnetic compression ratio $\geq 1.2$, density ratio $\geq 1.2$, and temperature ratio $\geq .83$.

For both IP shock property calculations and the ESP spectral fits, we use high-time resolution data to ensure that each time window average has valid data points. Our shock list is also fully representative of the six different classifications of ESP enhancements from Lario et al. (2003). A typical ESP event from the data set including particle enhancements, solar wind properties, and magnetic field is shown in Figure 1. Throughout the analysis, we do not exclude any complicated (e.g., time profile) events or offer any preferential treatment for well-defined shocks and their associated ESP events. However, we exclude 18 events where the ACE/SWEPAM data are not valid (missing) or sparse. The list of events is included in Table A1 in the Appendix. The detectors on ACE are now long past their originally planned lifetime, and with age the proton density and temperature observations became increasingly sparse and at times did not make good measurements (ACE Science Center)[6]. During these times, we found erroneous shock property calculations in various time windows and thus excluded these events from our analysis. We define fixed time intervals for the upstream and downstream time windows of 2, 4, 8, 12, 14, and 20 minutes and calculate the shock properties across the shock during these intervals. For the ESP spectral calculations, we examine the upstream and downstream regions and fit a single power-law form to particle fluxes at the lowest four energies from ACE/EPAM (0.47 to 3.21 MeV) using time windows of 2, 4, 8, 12, 14, 18, 20, 60, 180, 360, and 540 minutes.

### 3.2. IP Shock Property Calculations

Similar to Kilpua et al. (2015), we calculate the shock parameters; speed jump ($|\Delta V|$); shock speed ($V_{sh}$); Alfvénic Mach number ($M_A$); shock obliquity ($\theta_{B_n}$); the magnetic field, density, and temperature ratios ($R_B$, $R_N$, $R_T$) from the Rankine–Hugoniot relations (e.g., Viñas & Scudder 1986) using 64 s level-2 plasma data from SWEPAM, and 16 s magnetic field data from MAG. Since the magnetic coplanarity fails for shock obliquity when the magnetic field upstream is parallel or perpendicular (0° and 90°) to the shock front, we calculate the shock normal, $\hat{n}$; the angle between the upstream magnetic field and the normal of the shock, using the mixed-mode method (Abraham-Shrauner & Yun 1976), which uses both plasma and magnetic field data as shown in Equation (1):

$$\hat{n} = \frac{+(B_{down} - B_{up}) \times ((B_{down} - B_{up}) \times (V_{down} - V_{up}))}{|(B_{down} - B_{up}) \times ((B_{down} - B_{up}) \times (V_{down} - V_{up}))|},$$

(1)

where $B_{down}$ is the component of the magnetic field downstream; $B_{up}$ the component of the magnetic field upstream; and $V_{down}$ and $V_{up}$ are components of the solar wind velocity downstream and upstream, respectively. Downstream properties of shocked plasma are often characterized by the Alfvénic Mach number, $M_A$, given by

$$M_A = \frac{|V_{up} \cdot \hat{n} \pm V_{sh}|}{V_A^{up}},$$

(2)

where $V_A^{up}$ is the upstream Alfvén velocity,

$$V_A^{up} = \left\langle \frac{B}{\sqrt{\mu_0 N_p m_p}} \right\rangle.$$

(3)

The speed jump, $|\Delta V|$, is the speed difference between the downstream and upstream regions of the bulk solar wind

---

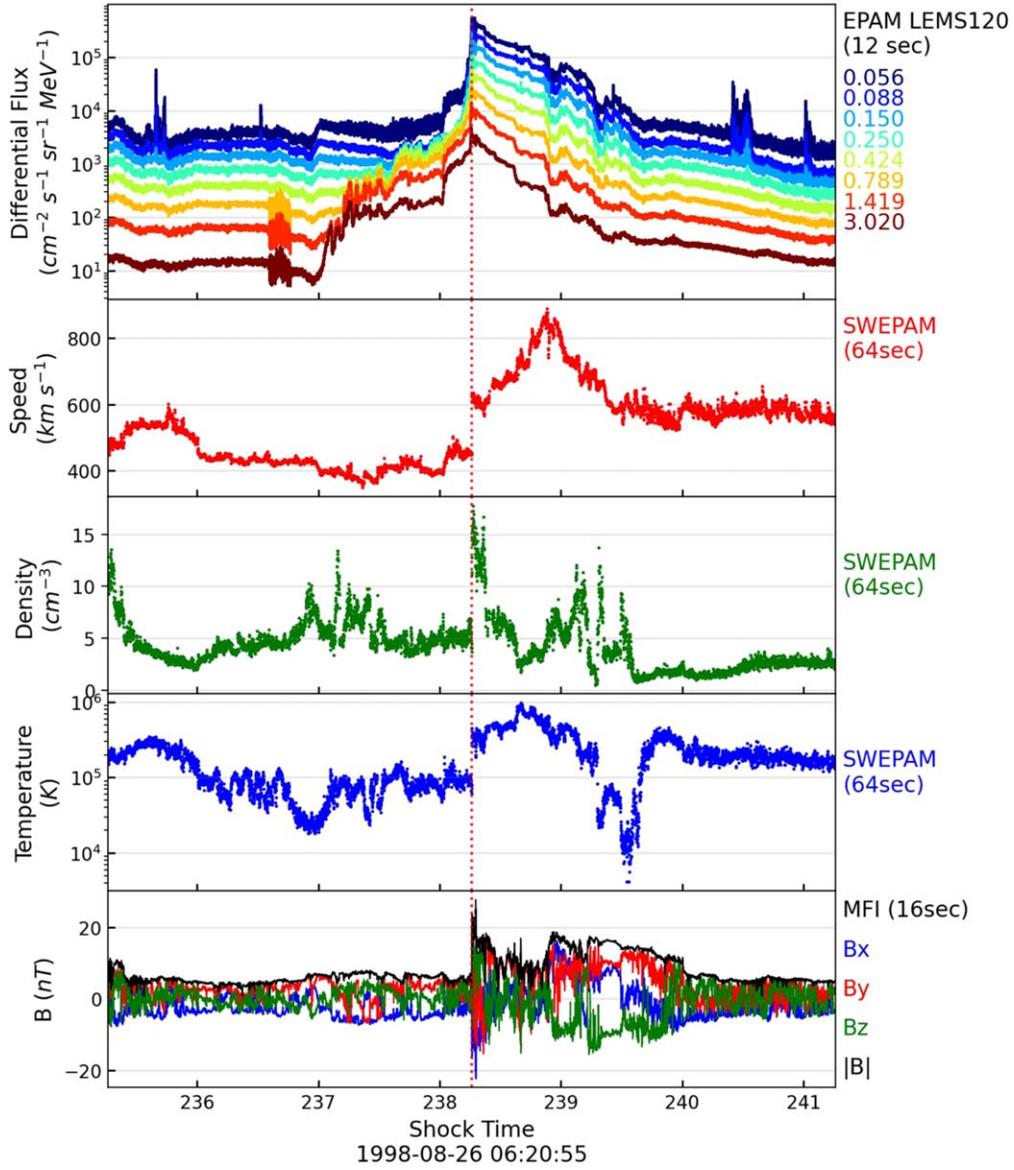

**Figure 1.** An example of particle and solar wind signatures during a typical Energetic Storm Particle (ESP) event. The top panel shows particle data from the LEMS120 instrument on ACE (∼.056 to ∼3.02 MeV). The bottom three panels show the magnitude of the solar wind velocity, density, and temperature from the SWEPAM instrument. The bottom panel shows the magnetic field vector (red, blue, and green lines) and magnitude (black line) from the MAG instrument. The vertical dotted red line shows the arrival time of the IP shock detected by the ACE spacecraft at 06:20:55 UT on 1998 August 26. This shock is driven by an interplanetary coronal mass ejection (ICME), with the sheath (turbulent magnetic field) and magnetic (field rotation) following.

velocity, as

$$|\Delta V| = |V_{\text{down}} - V_{\text{up}}|. \tag{4}$$

The shock speed, $V_{\text{sh}}$, is calculated in the solar wind frame of reference using the mass flux over the shock,

$$V_{\text{sh}} = \left| \frac{N_p^{\text{down}} V_{\text{down}} - N_p^{\text{up}} V_{\text{up}}}{N_p^{\text{down}} - N_p^{\text{up}}} \cdot \hat{n} \right|, \tag{5}$$

where $N_p^{\text{down}}$ is the proton density downstream and $N_p^{\text{up}}$ is the upstream proton density. The angle between the shock normal vector and the upstream magnetic field lines, the shock obliquity ($\theta_{B_n}$), is calculated as

$$\theta_{B_n} = \frac{180°}{\pi} \arccos\left( \frac{|B^{\text{up}} \cdot \hat{n}|}{\|B^{\text{up}}\| \|\hat{n}\|} \right). \tag{6}$$

Finally, the magnetic, density, and temperature ratios are the downstream mean value for the time window divided by the upstream mean value of the same time window. These ratios are given by

$$R_B = \frac{B_{\text{down}}}{B_{\text{up}}} \tag{7}$$

$$R_N = \frac{N_{\text{down}}}{N_{\text{up}}} \tag{8}$$

$$R_T = \frac{T_{\text{down}}}{T_{\text{up}}}. \tag{9}$$

Since the shock is a discontinuity and its properties tend to be localized, we objectively choose to use "inclusive time" windows of 2, 4, 8, 12, 14, and 20 minutes upstream and downstream (i.e., "across the shock") from the observed shock time. The 8 minute time window is chosen explicitly to cross-validate our calculations against the Helsinki database values in which a similar 8 minute sampling interval is used. The inclusive method selects data in sequential intervals that include the previous time-sampling data (shown in Figures 2(a) and (b)). For example, when calculating the values for the 8 minute time window (~8 data points), the data from the previous 4 minute (~4 data points) window is included in the calculation.

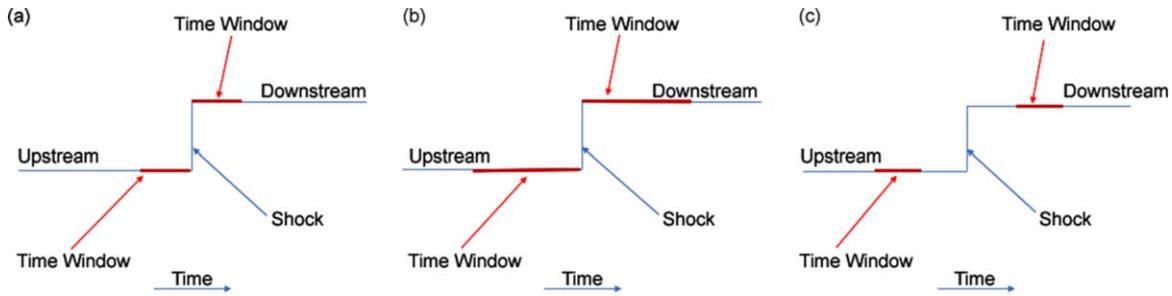

**Figure 2.** Time-sampling windows for the derived shock properties are done using an inclusive sampling method. Panel (a) shows the first time window where the data are taken from the time of the shock to the end of the time window; panel (b) shows the following time window, where the data are taken from the time of the shock to the end of that time window, including the previous selection. (c) Calculation of the ESP spectral indices is done using an exclusive time-sampling window, as illustrated. Here, data from the previous window are not included in the subsequent time interval calculations.

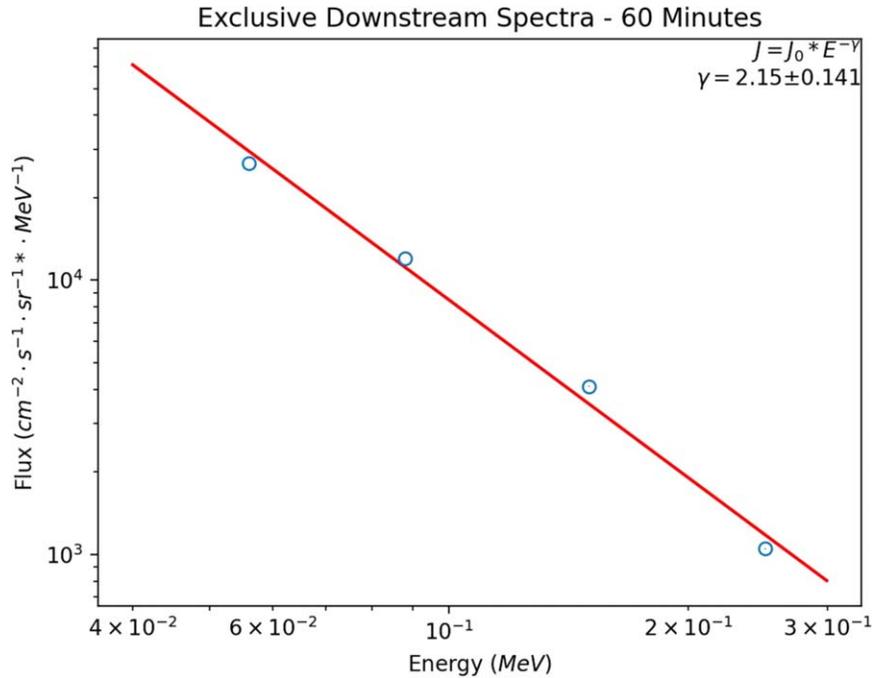

**Figure 3.** A sample power-law fit of the form $J(E) = J_o E^{-\gamma}$ for a selected time interval. To fit the spectra in all events, we use the lowest four energy channels, P1'–P4' (centered at .056 MeV, .088 MeV, .150 MeV, and .250 MeV), from the LEMS120 telescope of EPAM.

### 3.3. ESP Property Calculations

Using 12 s particle flux data from EPAM, we calculate the spectral index of energetic particles by fitting a power law of the form $J(E) = J_o E^{-\gamma}$ for the lowest four energy channels centered on .056 MeV, .088 MeV, .150 MeV, and .250 MeV; an example of the downstream spectral fit is shown in Figure 3. All fits are weighted by the propagated flux uncertainties. The fitting process is done for all time intervals using an "exclusive" time-sampling interval both upstream and downstream of the shock. Figures 2(a) and (c) show the exclusive method. For example, in the downstream 60 minute interval, the data from any time interval below 20 minutes is not included (only 40 minutes of data is used from shock plus 20 minutes to shock plus 60 minutes) in the spectral-index fit calculation. Spectral fits enable us to better understand the ensemble behavior of particles close to the shock and how they evolve away from the shock.

## 4. Results

### 4.1. Shock Property Distributions

Figure 4 shows the distribution of the IP shock properties calculated for the 2 minute (red) and 20 minute (blue) sampling intervals for all 278 events. We find the largest variation in values is for the shock speed jump ($|\Delta V|$), where the median value for the 2 minute time window is 39.41 and 53.44 km s$^{-1}$ for the 20 minute time window. Results show that the IP shocks cover all angles with a bias toward quasi-perpendicular events, which are observed more frequently. From 2 minutes to 20 minutes, the shock speed ($V_{sh}$) changes from a median value of 483.52 to 488.07 km s$^{-1}$. The Alfvénic Mach ($M_A$) number has median values of 2.16 and 2.23. The magnetic ($R_B$), density ($R_N$), and temperature ($R_T$) ratios all have median values

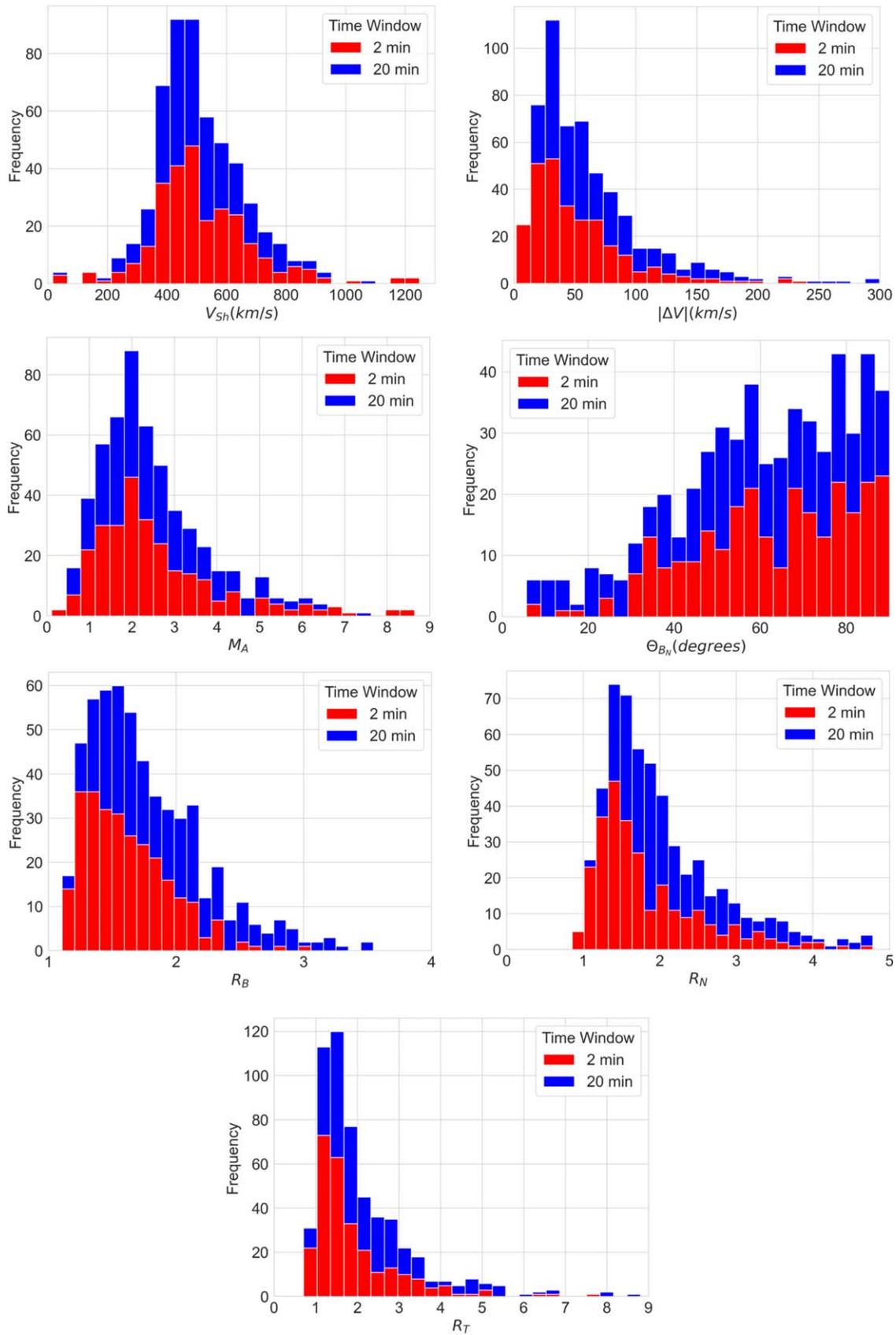

**Figure 4.** Distributions of all derived shock properties shown in stacked histograms for 2 minute (red) and 20 minute (blue) intervals for all 278 events.

between 1.5 and 2 for both the 2 minute and 20 minute intervals.

Since the shock property distribution is non-Gaussian for all time windows, we use the median absolute deviation (MAD) to quantify the statistical spreading behavior of the shock property. The MAD is given by MAD = median($|X_i - \overline{X}|$, where the $X_i$ is a shock property data point and $\overline{X}$ is the median of all the data points for that shock property. The MAD describes the variability of each examined property by quantifying how spread out the data are. It is also a robust measurement of dispersion that is less affected by outliers and is scale free. We also change MAD to a percentage value for easy comparisons among different derived properties (% = MAD*100/Median).

Box plots in Figure 5 provide a summary of each shock property and their variations. Corresponding median values for each time window are also shown. Outlier values are noted by the whiskers, and the MAD values (%) are shown for the respective time windows. Table 1 shows the median values for all derived IP shock properties over each time-sampling window, the MAD value, and the MAD as a percentage for all the time windows (2–20 minutes).

### 4.2. ESP Spectral Index Distributions

Using the derived indices for all events and for different time windows, we plot the distributions for both regions, upstream and downstream, around the shock and using the exclusive method described earlier. Figure 6 highlights the spectral index distributions from four different sampling windows; 2 minutes, 20 minutes, 60 minutes, and 540 minutes. Table 2 shows the median spectral indices for all time-sampling intervals. The spectral indices become more flat (i.e., harder) farther away from the shock or as the time windows becomes larger. This trend happens in both the upstream as well as downstream values. Similar to Reames (2012b) who noted that choosing a period of 12 or 16 hr upstream of the shock made little difference in the spectra for most shocks, we also find that selecting a period of above 540 minutes (9 hr) makes little difference in our spectral indices, especially in the downstream region of the shock. We note that it is possible the upstream spectra period contains accelerated particles from a previous event in cases where solar activity is high. In our investigation of 278 events, we observed that 15 of them, approximately 5% of the total, had a previous shock occurrence within a 540 minute window (+/−20 minutes). In the upstream region from the 2 minute to 540 minute interval, the spectral index value decreases from 1.76 to 1.63. The downstream region spectra index values drop slightly from 1.77 in the 2 minute time window to 1.74 in the 540 minute window. A similar trend occurs in the values upstream of the shock although the spectra is more flat (harder) in the longest time window of 540 minutes. The slight variation in spectral indices suggests that energetic particles within 540 minutes, both upstream and downstream of the IP shocks, are accelerated by the IP shocks.

MAD spectral index variations of 0.46 for all time windows in the downstream regime, corresponding to a change of 26.05%, and a slightly higher MAD value in the upstream regime of 0.52 or 30.53% change. Figure 7 summarizes the spectral index variations (time window colors match the histograms in Figure 6) and their median values as well as the MAD values as a percentage in bold. The whiskers portray the outliers in the calculations.

### 4.3. Upstream and Downstream Relations

We now examine the upstream-downstream relations by investigating how derived shock parameters evolve away from the shock in both the upstream and downstream directions. Figure 8 shows an epoch analysis for all events and for three parameters, namely, solar wind speed, magnetic field magnitude, and proton density. Here, we plot the median solar wind speed (red line) two days before and after the shock along with the upper (+ 25%) and lower quantiles (−25%) (black lines). As anticipated, epoch analysis shows that the solar wind velocity, magnetic ratio, and density ratio all show significant enhancements at the arrival time of the shock. The solar wind velocity jumps on average from ∼425 to ∼475 km s$^{-1}$, the average magnetic field jumps from ∼6 nT to ∼12 nT, and the density jumps from ∼5 n/cc to ∼11 n/cc. Note that the magnetic field and the density fall sharply away from the shock, while the solar wind speed is almost fixed away from the shock. For a typical event, a similar ratio for the density and magnetic field is consistent with a quasi-perpendicular shock. Note that to have a compression ratio of ∼2, the shock speed in the spacecraft frame should be about 525 km s$^{-1}$.

Next, we examine the evolution of shock properties when moving away from the shock for the three properties shown in Figure 9 using the Pearson correlation coefficient (*r*). Figure 9(a) shows the evolution of the correlation coefficient of shock parameters from 2 minutes up to 120 minutes in 2 minute increments. The correlation is determined using two methods: Case I, time windows move away from the shock in both the upstream and downstream directions in 2 minute steps (as in Figure 2(c)); and Case II, upstream time window is fixed at 2 minutes and the downstream time window moves away from the shock in 2 minute steps. The latter is done to investigate how far the downstream time windows can still be used to yield shock information. For both cases, we find a positive correlation that decreases as the time window moves away from the shock for all parameters.

For the moving-upstream and -downstream time windows (Case I), we find that the $|\Delta V|$ correlation falls from $r > 0.8$ to $r \sim 0.7$ within ∼18 minutes from the shock; it then plateaus at $r \sim 0.7$ for up to ∼70 minutes away from the shock and then falls slowly but monotonically to correlation values below 0.7 (blue arrow). In contrast to the $|\Delta V|$, $R_B$ and $R_N$ fall at a faster rate, reaching a correlation of ∼0.5 at ∼12 minutes and ∼30 minutes respectively. Here, the threshold is set at 0.5 because both quantities fall reasonably faster than that of the $|\Delta V|$. We emphasize that upstream could theoretically be any value as the shock has no information of what is ahead. However, this ensemble analysis on numerous events informs about the average time window that can be used on both sides of the shock and still maintain a reasonable shock parameter value.

For Case II where we investigate the evolution of these quantities using a fixed-upstream window of 2 minutes close to the shock and a moving-downstream region, results (Figure 9(b)) show that for the $|\Delta V|$, the evolution of $r$ is similar to that of Case I in a way that it falls slowly. However, the correlation is generally higher and falls below 0.7 after ∼110 minutes. The $R_B$ evolution is similar to Case I and drops below $r \sim 0.5$ after ∼26 minutes; it then plateaus and drops off slowly into weaker relations. In contrast, the $R_N$ here behaves differently from Case I as it falls linearly to $r \sim 0.5$ in ∼65 minutes with a correlation that is significantly higher than that of Case I.

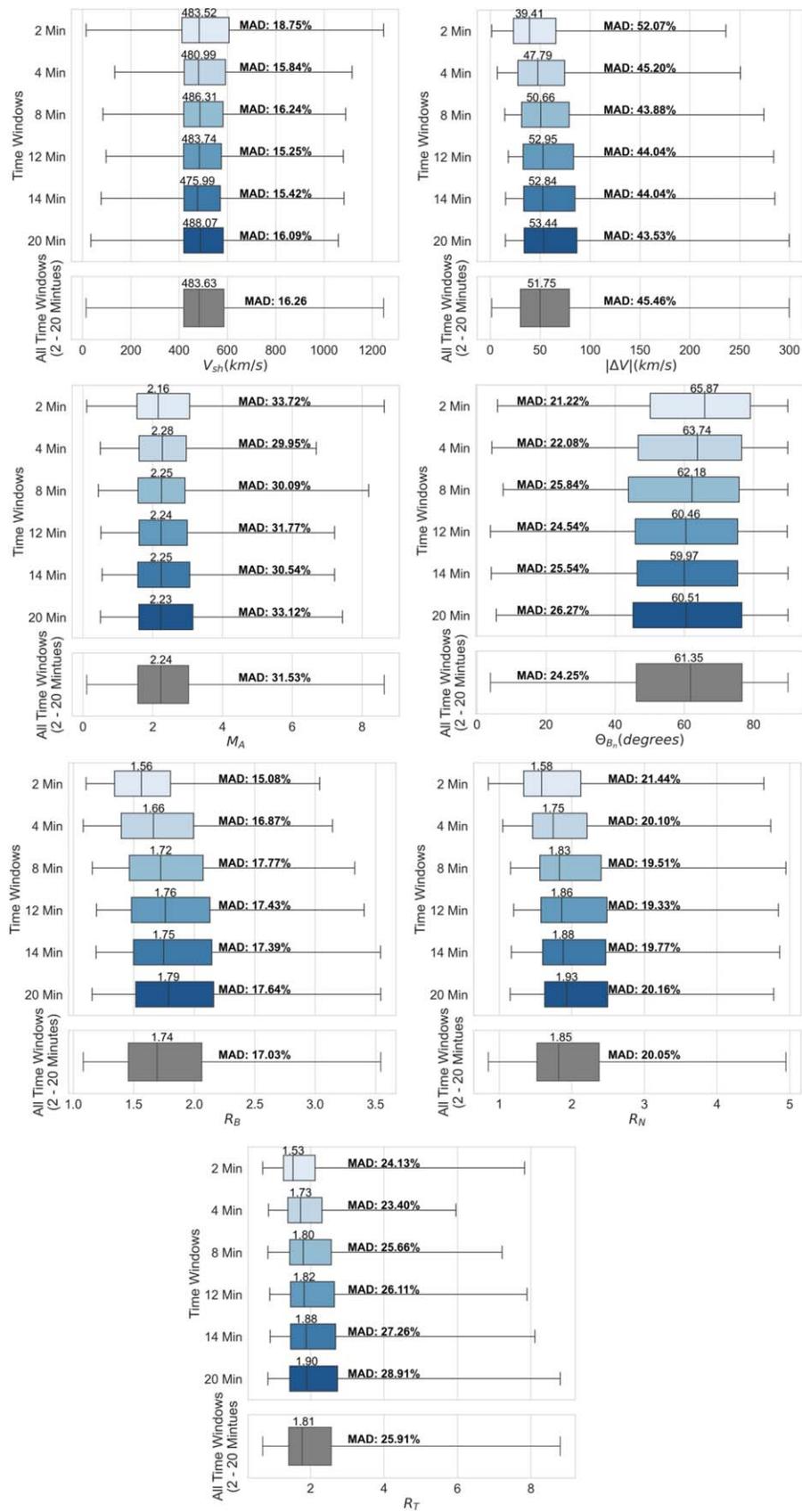

**Figure 5.** Box plots show the variation in each shock property. The time windows are show on the y-axis. The median value is shown above the box plot, the whiskers display outliers, and the MAD as a percent is displayed in bold for each time window. We note that there are several events in which the data around the shock time are missing or bad and cause erroneous values. These events have been removed from the data set (see Table A1 in Appendix), and a more detailed explanation of the erroneous data is given in Section 3.

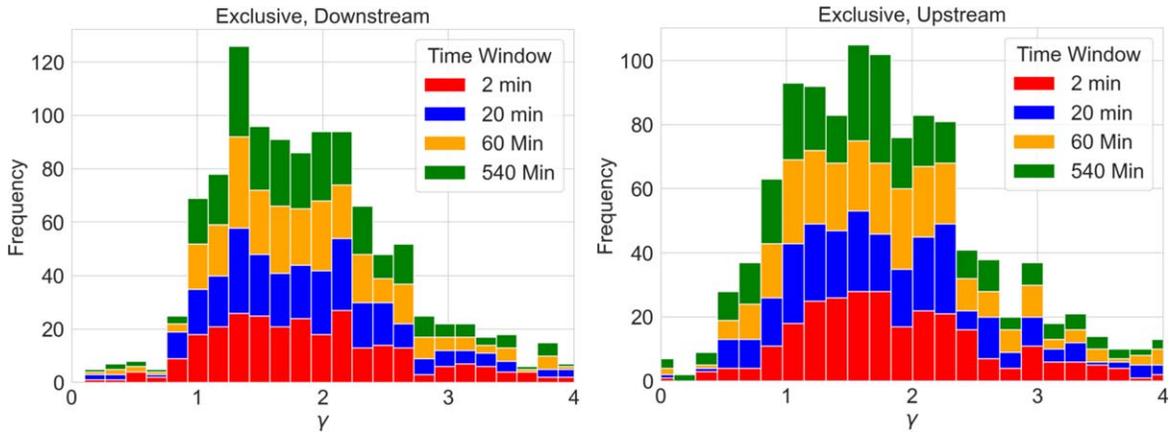

**Figure 6.** We show the spectral index ($\gamma$) distributions for four different time windows. The red bar indicates spectral indices for the 2 minute time window; blue for the 20 minute window; orange for the 60 minute time window; and green for the largest time window of 540 minutes. Panel (a) shows the index values calculated using the exclusive method downstream of the shock. Panel (b) shows the distribution of spectral index value using the exclusive method upstream of the shock.

**Table 1**
Calculated Shock Properties for each Time Window Interval

| Time Window Minutes | IP Shock Properties (278 Events) | | | | | | | | | | | |
|---|---|---|---|---|---|---|---|---|---|---|---|---|
| | $|\Delta V|$ | | | $V_{sh}$ | | | $M_A$ | | | $\theta_{B_n}$ (deg) | | |
| | | | (km s$^{-1}$) | | | | | | | | | |
| | Med | MAD | MAD % | Med | MAD | MAD % | Med | MAD | MAD % | Med | MAD | MAD % |
| 2 | 39.41 | 20.52 | 52.07% | 483.52 | 90.66 | 18.75% | 2.16 | 0.73 | 33.72% | 65.87 | 13.98 | 21.22% |
| 4 | 47.79 | 21.60 | 45.20% | 480.99 | 76.17 | 15.84% | 2.28 | 0.68 | 29.95% | 63.74 | 14.08 | 22.08% |
| 8 | 50.66 | 22.23 | 43.88% | 486.31 | 78.96 | 16.24% | 2.25 | 0.68 | 30.09% | 62.18 | 16.06 | 25.84% |
| 12 | 52.95 | 23.32 | 44.04% | 483.74 | 73.79 | 15.25% | 2.24 | 0.68 | 31.77% | 60.46 | 14.84 | 24.54% |
| 14 | 52.84 | 23.27 | 44.04% | 475.99 | 73.41 | 15.42% | 2.25 | 0.71 | 30.54% | 59.97 | 15.31 | 25.54% |
| 20 | 53.44 | 23.26 | 43.53% | 488.07 | 78.53 | 16.09% | 2.23 | 0.74 | 33.12% | 60.51 | 15.90 | 26.27% |
| (2–20) | 51.75 | 22.33 | 45.46% | 483.63 | 78.41 | 16.26% | 2.24 | 0.71 | 31.53% | 61.35 | 15.08 | 24.25% |

| | $R_B$ | | | $R_N$ | | | $R_T$ | | |
|---|---|---|---|---|---|---|---|---|---|
| | Med | MAD | MAD % | Med | MAD | MAD % | Med | MAD | MAD % |
| 2 | 1.56 | 0.24 | 15.08% | 1.56 | 0.34 | 21.44% | 1.53 | 0.37 | 24.13% |
| 4 | 1.66 | 0.28 | 16.87% | 1.75 | 0.35 | 20.10% | 1.73 | 0.41 | 23.40% |
| 8 | 1.72 | 0.31 | 17.77% | 1.83 | 0.36 | 19.51% | 1.80 | 0.46 | 25.86% |
| 12 | 1.76 | 0.31 | 17.43% | 1.86 | 0.36 | 19.33% | 1.82 | 0.48 | 26.11% |
| 14 | 1.75 | 0.3 | 17.39% | 1.88 | 0.36 | 19.77% | 1.88 | 0.51 | 27.26% |
| 20 | 1.79 | 0.32 | 17.64% | 1.93 | 0.39 | 20.16% | 1.90 | 0.55 | 28.91% |
| (2–20) | 1.74 | 0.29 | 17.03% | 1.85 | 0.37 | 20.05% | 1.81 | 0.46 | 25.91% |

**Table 2**
ESP Spectral Indices (278 events)

| Time window (minutes) | Downstream | | | | Upstream | | | |
|---|---|---|---|---|---|---|---|---|
| | $\gamma$ (Med) | MAD | MAD % | $\sigma$ (Std Dev) | $\gamma$ (Med) | MAD | MAD % | $\sigma$ (Std Dev) |
| 2 | 1.77 | 0.48 | 26.95 | 0.78 | 1.76 | 0.48 | 27.11% | 0.79 |
| 4 | 1.81 | 0.49 | 27.20% | 0.77 | 1.72 | 0.49 | 28.66% | 0.81 |
| 8 | 1.81 | 0.50 | 27.64% | 0.76 | 1.72 | 0.54 | 31.28% | 0.82 |
| 12 | 1.81 | 0.50 | 27.50% | 0.75 | 1.72 | 0.5 | 28.88% | 0.81 |
| 14 | 1.84 | 0.50 | 26.90% | 0.75 | 1.70 | 0.51 | 30.03% | 0.81 |
| 18 | 1.82 | 0.50 | 27.23% | 0.76 | 1.71 | 0.51 | 30.15% | 0.81 |
| 20 | 1.8 | 0.48 | 26.47% | 0.75 | 1.70 | 0.53 | 31.31% | 0.8 |
| 60 | 1.83 | 0.50 | 27.41% | 0.74 | 1.70 | 0.52 | 30.60% | 0.81 |
| 180 | 1.77 | 0.44 | 24.86% | 0.73 | 1.66 | 0.53 | 31.74% | 0.81 |
| 360 | 1.75 | 0.42 | 24.11% | 0.72 | 1.66 | 0.53 | 31.70% | 0.85 |
| 540 | 1.74 | 0.41 | 23.37% | 0.72 | 1.63 | 0.54 | 33.08% | 0.83 |
| MAD (2–540) | 1.79 | 0.46 | 26.05% | 0.75 | 1.7 | 0.52 | 30.53% | 0.82 |

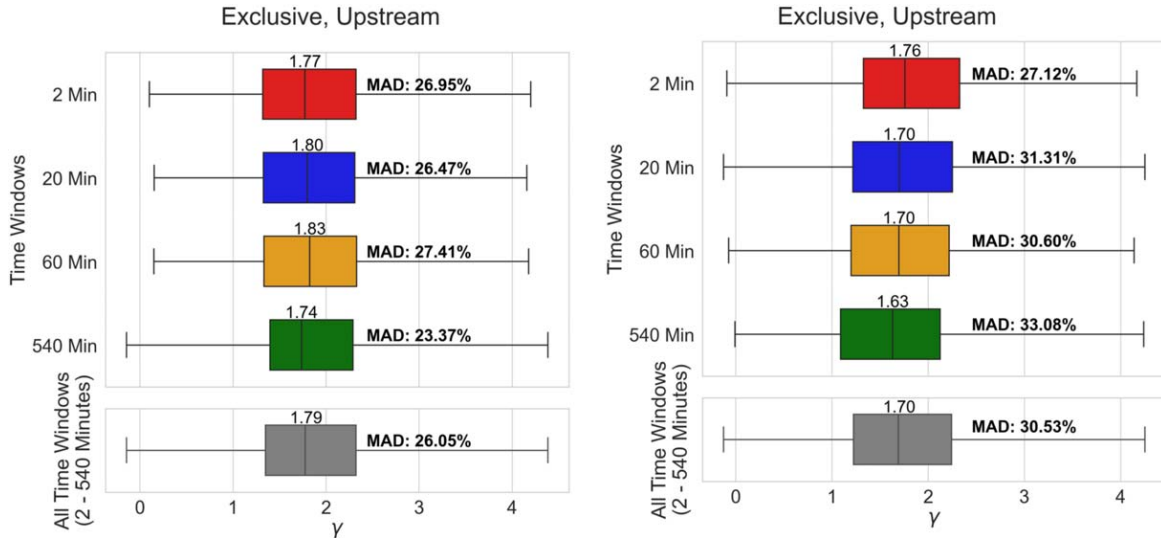

**Figure 7.** Box plots for the upstream and downstream spectral indices along with the median values above each box plot and the MAD as a percentage in bold to the right. The individual time window colors match the previous distribution plots in Figure 6. For all time windows (gray-colored box plots) we find that the index values vary more from their median in the upstream regime of the shock.

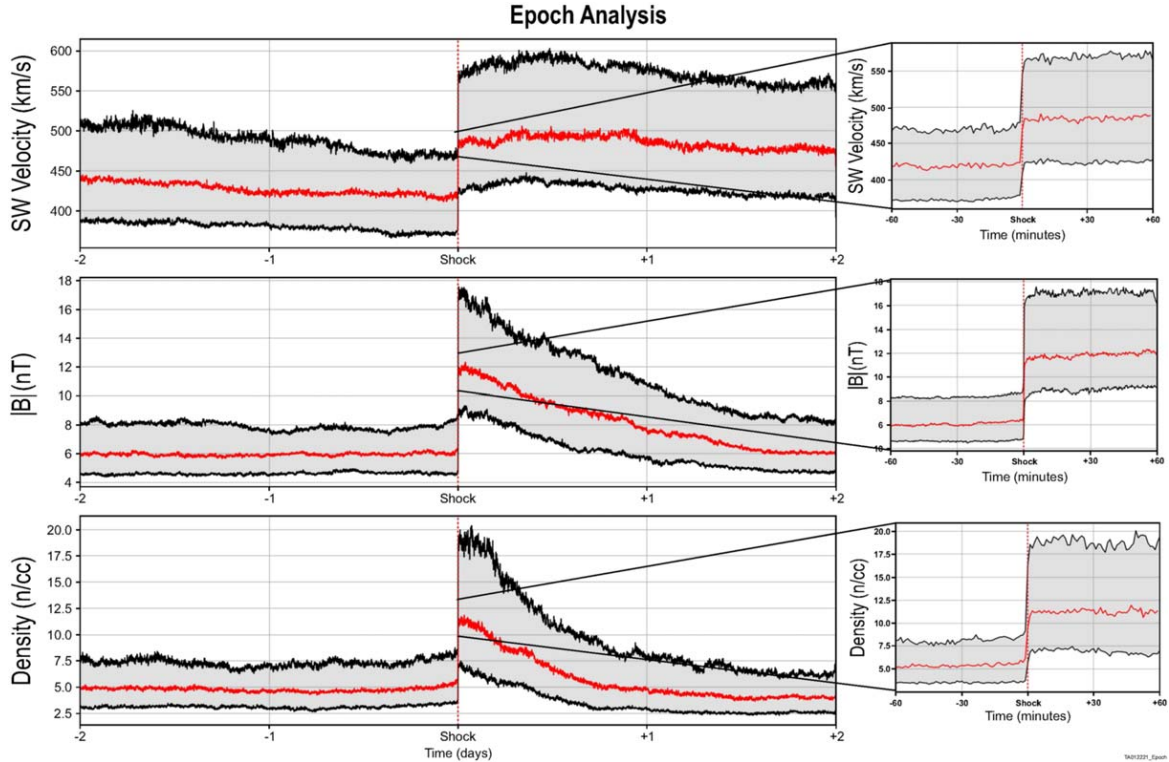

**Figure 8.** Averaged solar wind parameters plotted 2 days before and after the shock for all 278 events. The middle red line represents the mean, with the upper and lower quantiles plotted in black. The zoomed-in plots show the 60 minutes before and after the shock in closer detail.

Overall, these results show that a sufficient time window is dependent upon the property being analyzed. The $|\Delta V|$ window is relatively constant for ∼120 minutes, $R_B$ is good for ∼30 minutes, and $R_N$ window is good for ∼10 minutes upstream and ∼60 minutes downstream. Beyond these windows, the upstream and downstream regions are less representative of the associated shock properties. Thus, the time windows should be constrained to within the derived $\Delta t$ above each property's correlation plateaus notated in Figure 9 by color-coded arrows.

## 5. Conclusions

We study the variations of the IP shock parameters and the associated ESP properties in 278 events derived using different time windows upstream and downstream away from the shock. Variability is expressed in terms of the MAD statistic. We find:

1. As anticipated, the most minimal deviations from the median are observed within the shortest time-sampling window of 2 minutes, which holds true for five out of the seven shock properties. In the $|\Delta V|$ we find that the

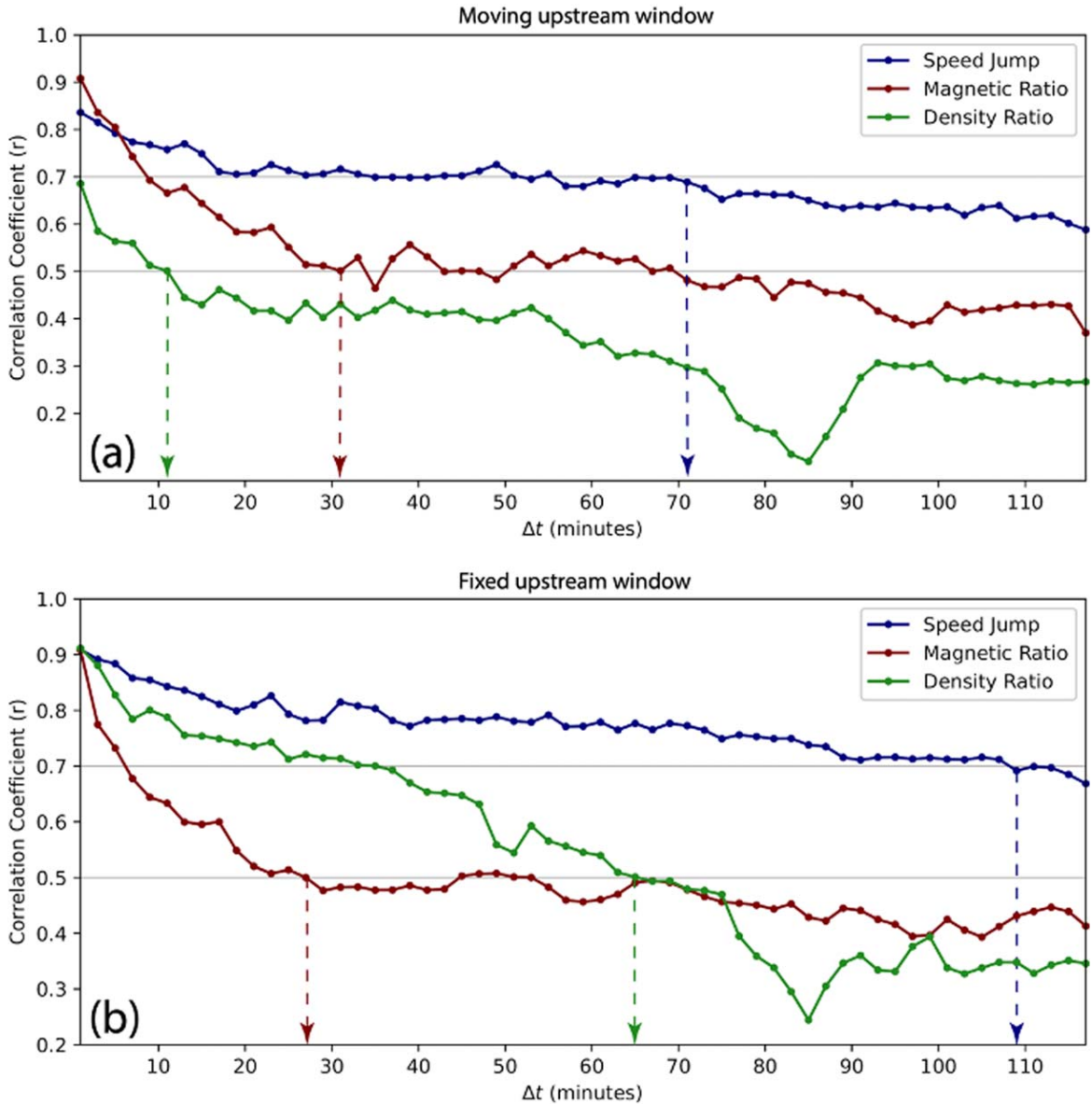

**Figure 9.** Evolution of the Pearson correlation coefficient ($r$) for three shock properties: $|\Delta V|$, $R_B$, and $R_N$. Each panel shows the evolution of the correlation over 2 hr period in 2 minute time window increments. Panel (a) uses a moving time window method across the shock (Case I); panel (b) uses a fixed 2 minute upstream time window and moving-downstream time window (Case II). The 0.5 and 0.7 thresholds are marked by horizontal lines. The time where the properties fall below the thresholds are noted by color-coded arrows.

14 minute time window has the smallest MAD of 73.41 km s$^{-1}$ from its median value of 475.99 km s$^{-1}$. We find the $M_A$ varies from the median by 0.68 in the 4, 8, and 12 minute time windows. Of the three ratios the temperature ratio, $R_T$, has the greatest variation from the median value, and the magnetic ratio, $R_B$, varies the least.

2. For ESP properties, the median spectral index value upstream of the shock decreases from 1.76 to 1.63 from the 2 minute to 540 minute time windows in a monotonic trend but with an MAD value of ~30%, indicating consistent and significant fluctuations of this order within each time window (Table 2). Downstream of the shock, the mean spectral index does not depend on the window interval strongly. The averaged spectral index is 1.79 with an MAD of ~26% (Table 2).

3. Determining a sufficient time window for shock property calculations is dependent on the property being analyzed.

Moving away from the shock in both upstream and downstream directions, the $R_N$ time window is good to ~10 minutes, $R_B$ is good to ~30 minutes, and the shock $|\Delta V|$ is good up to 60 minutes. On the other hand, fixing the upstream window close to the shock and moving only in the downstream direction, then the $R_N$ time window holds up to ~60 minutes downstream, $R_B$ window up to ~30 minutes, and the $|\Delta V|$ up to ~110 minutes. Beyond these windows, we suggest that derived shock properties are less representative of the shock properties.

Table 2 reports uncertainties corresponding to the standard deviation ($\sigma$) from all events for each time window. We opted to use this statistic instead of the the fitting uncertainty because the latter was too small due to the very large fluxes, which inherently have small errors ($<1\%$). $\sigma$ in this case better represents the spread of the data points for each time window. Interestingly, small and large time windows have very comparable deviations. At large time window values, event selections may extend to regions that are not related to the examined shock or the ESP event, including mixing with particles from previous events for the latter case. While our selection of 20 minutes for the shock properties and 540 minutes for the ESP events limits this scenario, it does not fit all events. Nonetheless, this does not seem to affect the final results that are based on a larger number of events.

For small time windows, the large fluctuation in the MAD could be related to several sources, and this variation should be anticipated when dealing with statistical studies, regardless of the purpose of the science study. For ESPs, the downstream region is observed to have a reservoir-like population of particles that are locally trapped that form similar intensity behaviors and thus a fixed spectral shape (e.g., Roelof et al. 1992; Reames 2012a). However, variations observed are still of a considerable value. The same applies for the upstream region; although temporal particle profiles closer to the shock are similar, large variations in the spectral indices are present (e.g., Desai et al. 2003; Santa Fe Dueñas et al. 2022). In MHD simulations, shocks moving through large-scale turbulence have been shown to have an energy spectrum independent of location along the shock (e.g., Giacalone and Nuegebuer 2008; Zank et al. 2010). Some IP shock events show no increase in particle intensity at the shock passage (e.g., Lario et al. 2003; van Nes et al. 1984). The lack of particle enhancement in these shocks can affect the spectral indices and therefore their derived variability coefficient.

Ambient variation of, for example, solar wind density and speed and local turbulent magnetic field, can affect shock geometry, compression ratio, and Mach numbers. These ambient variations inevitably contribute to the observed deviations examined in this study. Note that these variations may have an implication on large SEP events as well and has been a subject of recent numerical simulations of large SEP events, e.g., Li et al. (2021).

For both ESP and IP shock parameters, we note that the data come from single-point measurements and there is always the possibility that the local region being sampled does not necessarily apply for the entire shock structure or ESP profile (e.g., Giacalone 2012).

This study quantifies the effect of the time-sampling window on derived IP shock and associated ESP properties. Variations are listed in Table 1 for the shock properties and Table 2 for the ESP properties. Results show an averaged MAD variability of 25.78% for shock properties within a 20 minute window and 28.29% for ESPs within a 540 minute window. These values provide quantitative constraints for IP shock modeling and ESP forecasting applications.

## Acknowledgments

This work is funded by NASA Grants O2R 80NSSC20K0290 and LWS 80NSSC19K0079. We acknowledge using data from the Helsinki Database of Heliospheric Shock Waves generated and maintained at the University of Helsinki http://ipshocks.fi/. The ACE SWEPAM solar wind and magnetic data are available for download from the Coordinated Data Analysis Web (CDAWeb) at https://cdaweb.gsfc.nasa.gov/index.html/ as is the particle flux data from EPAM. K.M. thanks Samuel Hart for his commitment to reviewing the paper and insightful questions. Support from NASA grants 80NSSC19K0075 and 80NSSC19K0831 at UAH is acknowledged. Author A.F. would like to acknowledge the support provided by the Deanship of Scientific Research (DSR) at the King Fahd University of Petroleum and Minerals (KFUPM) for funding his work through project No. DF181010.

## Appendix

This appendix includes Table A1, a list of IP shock times that are excluded from our data analysis. See Section 3.1 for details.

**Table A1**
Excluded Shock Events

| | |
|---|---|
| 1998-04-30 08:43:33 | 2002-03-18 12:37:09 |
| 1998-07-05 03:16:14 | 2002-09-07 16:08:45 |
| 1999-09-12 03:21:12 | 2002-12-22 12:16:31 |
| 2000-04-06 16:03:59 | 2003-05-29 18:30:18 |
| 2000-09-17 16:57:15 | 2004-07-26 22:27:43 |
| 2001-03-31 00:22:42 | 2004-11-09 09:14:08 |
| 2001-03-31 21:39:30 | 2005-09-09 13:16:53 |
| 2001-04-08 10:32:50 | 2005-09-15 08:29:40 |
| 2001-04-13 07:05:38 | 2013-05-24 17:36:20 |

## ORCID iDs

K. Moreland  https://orcid.org/0000-0002-6202-8565
M. A. Dayeh  https://orcid.org/0000-0001-9323-1200
G. Li  https://orcid.org/0000-0003-4695-8866
A. Farahat  https://orcid.org/0000-0002-0022-5782
R. W. Ebert  https://orcid.org/0000-0002-2504-4320
M. I. Desai  https://orcid.org/0000-0002-7318-6008